\documentclass[aps,nopacs,nofootinbib,twocolumn]{revtex4}
\usepackage{epsf,latexsym}
\usepackage{amsfonts}
\usepackage{graphicx}
\usepackage{trackchanges}

\begin{document}

\title{Conditions for graviton emission in the recombination
of a delocalized mass}
\author{Alessandro Pesci\footnotetext{e-mail: pesci@bo.infn.it}}
%\author{Alessandro Pesci}
%\email{pesci@bo.infn.it}
\affiliation
{INFN Bologna, Via Irnerio 46, I-40126 Bologna, Italy}

\begin{abstract}
In a known gedanken experiment, a delocalized mass is recombined while the 
gravitational field sourced by it is probed by another (distant) particle;
%%proof0 
in it, this is used to explore a possible tension 
%%proof1
between complementarity and causality in case the gravitational field 
entangles with the superposed locations, a proposed resolution being graviton 
emission from quadrupole moments. Here, we focus on the delocalized particle 
(forgetting about the probe and the gedanken experiment) and explore the 
conditions (in terms of mass, separation, and recombination time) for graviton 
emission. Through this, we find that the variations of quadrupole moments in 
the recombination are generically greatly enhanced if the field is entangled 
compared to if it is sourced instead by the energy momentum expectation value 
on the delocalized state (moment variation $\sim m \, d^2$ in the latter case, 
with $m$ mass, $d$ separation). In addition, we obtain the (upper) limit 
recombination time for graviton emission growing as $m$ in place of the naive 
expectation $\sqrt{m}$. In this, the Planck mass acts as threshold mass 
(huge, for delocalized objects): no graviton emission is possible below it, 
however fast the recombination occurs. If this is compared with the decay 
times foreseen in the collapse models of Di{\' o}si and Penrose (in their 
basic form), one finds that no (quadrupole) graviton emission from 
recombination is possible in them. Indeed, right when $m$ becomes large enough 
to allow for emission, it also becomes too large for the superposition 
to survive collapse long enough to recombine. 
\end{abstract}

\maketitle

$ $
%%%%%%%%%%%%%%%%%%%%%%%%%%%%%%%%%%%%%%%%%%%%%%%%%%%%%%%%%%%%%
\section{Introduction and background}

To date, there is still no direct evidence for a nonclassical nature 
of the gravitational field. Quantum effects accompanying gravity are expected 
to unavoidably show up at the Planck length scale $l_p$. Many of the proposed 
tests on quantumness of gravity  involve consideration of cosmological or
astrophysical circumstances, in which the cumulative effects over long 
distances might compensate for the smallness of $l_p$. A trouble with this 
is the lack of full control of the experimental circumstances, i.e., 
our degree of ignorance/uncertainty concerning the model of the universe, 
the source, and the propagation of the signal to the observer.

The alternative is laboratory tests on systems suitably designed to let 
potential quantum features of gravity to show up, following a proposal 
proposed originally by Feynman. The idea is \cite{Feynman} that the final 
quantum state of a system in which a delocalized mass is allowed to 
gravitationally interact with another mass ought to be different depending 
on whether the mediating field is quantum or classical.

The difference appears very hard to detect, but advances in quantum 
technologies have, by now, made these kind of tests feasible, or at least 
conceivable in practice. Quantum systems are used
as sources of the field typically in a superposition 
of locations (\cite{CarnA, Hug} for review and discussion), 
the main difficulty to face in this kind of effort being 
decoherence~\cite{Aspelmeyer}.

Suggestions have been made, for example, to look at stochastic fluctuations 
of quantum origin in the gravitational 
field \cite{Bassi1, Bassi2, Anastopoulos}. 
Starting from \cite{Bose, Marletto}, a new twist has been given 
to the subject with the proposal to directly check quantum coherence aspects 
of gravity, in the form of the ability of the gravitational field to entangle 
systems initially prepared in a separable state. The point is that 
no entanglement can be created by two parties that communicate exclusively 
through a certain local channel if the latter 
is classical \cite{NielsenChuang, Hor4}; the appearance of entanglement 
between two accessible parties
from initial conditions of no entanglement would then exhibit a nonclassical 
nature for the mediating unaccessed channel \cite{KZPP}. 

Strictly speaking, we can argue that the communication processes involving 
the gravitational field
might be nonlocal, yet causal \cite{FragkosKoppPikovski}. 
If this is the case, there would be no mediators and the just-mentioned 
creation of entanglement would prove that quantum sources do create 
superpositions of geometries, yet without gravitons to mediate 
this \cite{FragkosKoppPikovski}. In this paper, we assume the locality 
of the gravitational channel, 
since our focus is on the possible emission of (physical) gravitons 
in the recombination of a delocalized source.
%%1

The proposals \cite{Bose, Marletto} have suggested to consider two masses, 
each one delocalized, interacting exclusively
through their gravitational field.
The masses are prepared in a separable state,
allowed to interact gravitationally,
and are eventually tested for entanglement.
The experimental requirements accompanying these kind of tests 
place their feasibility in a hopefully not-so-far future. 
This possibility appears even closer
when looking at \cite{Plenio}.
In it,
an experimental setup is considered
in which the strength of the gravitational interaction is increased
through use of a very heavy (not delocalized) mass, which acts as a mediator
between an unlocalized mass and an ancillary qubit.

\begin{figure}[h!]
  \includegraphics[width =8cm]{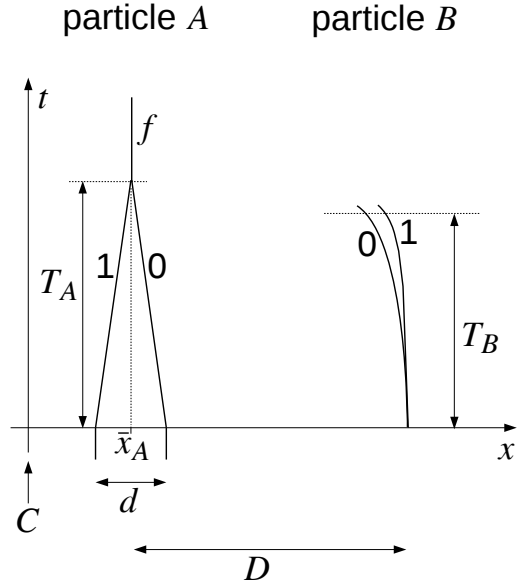}
  \caption{Setup of the thought experiment 
           \cite{MariDePalmaGiovannetti, Belenchia1} 
           as used in the analysis here (see main text). 
           The $x$ coordinate is distance taken from Alice's system's 
           center of mass $C$ (lab + her particle $A$), 
           and $C$'s worldline acts as the time axis.
           At a same time ($t = 0$), Alice starts recombining $A$, 
           from a (held from long before) superposition of locations 
           (with separation $d$), 
           and Bob releases his particle $B$ 
           located at a distance $D \gg d$ from $A$.
           Alice completes her task in a time $T_A$
           while Bob checks the position of $B$ at $t = T_B$.
           The labels 0 and 1 tag the superposed configurations of
           the system (no superposition for particle $B$ 
           in case gravity is not able
           to entangle). $f$ tags $A$ when it is undelocalized, 
           assuming it is located at a small distance $\bar x_A$ from $C$.}
  \label{fig}
\end{figure}

Building on
Feynman's observation,
other circumstances can be considered, 
in which,
even 
leaving the actual feasibility apart,  
the difference between 
the effects of quantum versus classical 
mediating gravitational field
can be evident and possibly 
rich in consequences at the theoretical level. 
One example, which is our starting point here,
is the configuration described schematically in Figure \ref{fig} 
\cite{MariDePalmaGiovannetti, Belenchia1}.
In it, a particle,
that we call Alice's particle $A$ with mass $m_A$,
is held (from a distant past) in a superposition of locations
(paths $0$ and $1$ with separation $d$),
and another particle, $B$ at a distance $D$, 
Bob's particle with mass $m_B$, 
(only) gravitationally interacts with $A$.
At a preassigned time,
Alice starts recombining $A$
and Bob releases $B$ (or decides not to).
Alice will perform her task in a time $T_A$,
and Bob will check for the position of $B$ after
a time $T_B$ from the release  
(we assume the experiment is local,
with Alice and Bob having no relative motion 
and sharing a local frame).
If the gravitational field indeed entangles,
the superposed positions of $A$ are accompanied by
different fields at $B$ 
%%change0
%(the two locations of $A$ give rise to different
%quadrupole moments for Alice's system),
%
(the two locations of $A$ give rise to different
quadrupole moments for Alice's system, 
and the gravitational
field at $B$, entangled with $A$'s locations,
is in a superposition of the two
states sourced by the corresponding quadrupole moments),
%%change1
and Bob can in principle be able, after a certain minimum time $T_{\rm wp}$,
to discriminate between the paths of $A$.

The configuration we are describing
was proposed 
in \cite{MariDePalmaGiovannetti}
(which elaborated on a previous investigation on gravitational tagging
of the path \cite{BaymOzawa})    
then reconsidered in~\cite{Belenchia1},
and further discussed 
in \cite{Belenchia2, Rydving, Grossardt, Danielson}.
In these works,
this is viewed similar to the scene of a gedanken experiment, 
the focus being on describing
a seemingly paradoxical situation
arising from requiring both the 
complementarity principle---meant as the fact that 
obtaining which-path
%%proof0 
as performed by Bob
%%proof1
must be incompatible with Alice being able 
%%proof0
to recombine coherently---and causality. 
In particular,
%%proof1
the perspective in \cite{MariDePalmaGiovannetti}
is to extract from the avoidance of a potential paradox
the existence of a minimum time Alice needs
in order to find if the state of $A$ is a coherent superposition
or a mixture.

The premise for the arising of a paradox
is, as mentioned, the assumption
that the gravitational field at Bob's location
can possibly allow for discrimination of the path of $A$.
If this is the case,
and if circumstances are such that the distance $D$
between $A$ and $B$ is larger than $T_A, T_B$
(we use Planck units through all the paper
unless explicitly stated otherwise),
%%proof0
then the which-path Bob performs 
apparently leads by complementarity principle to superluminal transmission
%%proof1
of information from Bob to Alice 
($A$ has to lose coherence).
If, on the other hand, the gravitational field at $B$ 
cannot distinguish the path, 
as would be the case if the field
is sourced by a mixture of the two paths, 
then no paradox at all can arise
(cf. \cite{Grossardt}).
The latter is, for example, the case if the
gravitational field at $B$ is sourced by
the expectation value $\langle T_{ab} \rangle$
of the energy--momentum tensor of $A$ (and its lab)
(this would be gravity in its semiclassical description,
matter is quantum but the gravitational field is classical):
in this case, 
the gravitational field feels a mixed state of paths,
and the positions of $B$ are not entangled
with the single possible paths.
This makes it clear that the assumption 
of the gravitational field being able to entangle,
that is (with locality assumption), 
of being quantum, 
is at the origin of the possible paradox.

According to \cite{Belenchia1, Belenchia2, Danielson},
the overcoming of the paradox is in the interplay
%%proof0 
between the spatial resolution, unavoidably finite, of Bob 
in determining the
position of $B$ (ideally the Planck length $l_p$), 
%%proof1 
and the fact that when 
Alice recombines $A$ quickly enough,
Alice's system   
emits gravitational radiation
(from the variation in the quadrupole
moment of Alice's system)
in the form of a quantum of radiation,
namely, a graviton.
In practice,
were circumstances 
(read, the difference of quadrupole moment of Alice's system
for the two positions of $A$)
such that Bob would be able to obtain which-path with $T_{\rm wp} < D$,
then, in case $T_A < D$,
$A$ would necessarily be above the threshold for graviton emission.
That is, the coherence of $A$ becomes 
destroyed regardless of what
Bob actually does, since we see that happens
even if Bob decides not to release the particle. 
In this, they then move one step farther with respect 
to \cite{MariDePalmaGiovannetti},
in that they do not only recognize the existence of a limit time  
in performing coherently the recombination, 
but also they identify the underlying reason for it.  
The absence of a paradox comes, then, as a consequence.

Here is the point of the present investigation.
The emission of quadrupole radiation is clearly conceivable
only if the quadrupole moment of Alice's system changes
in the recombination.
This is precisely what one would expect
if the gravitational field has values 
entangled
with the superposed positions of $A$.
One might then think of the emission of radiation by Alice's system
during recombination of $A$ 
as a way to tag the ability of the gravitational field to become 
entangled with the path,
regardless of any possible recourse to a test particle $B$.
In the case of the classical gravitational field, 
sourced by the expectation value of the energy--momentum tensor on the
delocalized state, we also expect a variation of quadrupole moment
in the recombination. 
This variation, however, turns out to be
generically negligibly smaller than the above, as we will see.
This aside,
the emission of (classical) radiation would originate from
the delocalized state as a whole and we would not expect it to affect
the coherence of $A$.  

In principle, we could then think of an experiment
in which the quantum nature of
the gravitational field might be checked,
under locality assumption,  
using only one delocalized mass: Alice's particle $A$ here, 
looking at that when it is recombined quickly enough---in a time below a certain threshold $T_A < T_{\rm emit}$---it emits a graviton, this being witnessed
by the abrupt loss of coherence
in an ideal situation in which the environmental influence on 
$A$ is taken under full control.    
One can guess that an experiment of this kind
is similar to an impossible task.
However, leaving any actual feasibility aside,
there might be, from a theoretical point of view,
an interest in having a closer look
at the conditions one has to require to allow for
graviton emission, and this is the aim of the present work. 
As a byproduct,
some indications on the (im)practicability of such an experimental scheme
will also emerge, as we will see
(as well as some specifications about the actual reasons behind
the avoidance of the paradox).

%%%%%%%%%%%%%%%%%%%%%%%%%%%%%%%%%%%%%%%%%%%%%

\section{Conditions for graviton emission}

%%0
Let us start from the analysis in \cite{Belenchia1, Belenchia2, Danielson},
which is within the just-discussed approximations and limits.
%%1
%
With reference to Figure \ref{fig}, it is shown that
in case the gravitational field felt by $B$
is entangled with the path of $A$,
then, assuming the spatial resolution is limited by the Planck length $l_p$,
%%proof0
Bob cannot perform which-path in a time $T_B < T_{\rm wp}$
with 
%%proof1
% 
\begin{eqnarray}\label{emit55.5}
T_{\rm wp} \sim \frac{D}{\sqrt{Q_A}} \, \,  D;
\end{eqnarray}
On the other hand,
during recombination of $A$, 
Alice's system will emit (at least) a graviton
if $T_A < T_{\rm emit}$ 
with
\begin{eqnarray}\label{emit54.2}
T_{\rm emit} \sim \sqrt{Q_A}.
\end{eqnarray}
In
 these equations, $Q_A$ is assumed to be
the order of magnitude of both the difference (taken positive)
between the quadrupole moments of Alice's system
for configurations 0 and 1
and for before and after recombination of $A$
(at leading order we have a quadrupole term, not dipole,
since the dipole contribution is suppressed by
momentum conservation of Alice's system \cite{Belenchia1}). 
From these results, we see that
whenever Bob can actually perform which-path in $T_B < D$,
this from (\ref{emit55.5}) means that we must have $D < \sqrt{Q_A}$,
and then if $T_A < D$, necessarily $T_A < \sqrt{Q_A}$, and Alice's system
emits \cite{Belenchia1, Belenchia2, Danielson}. 
No paradox can then arise.

%%0
The actual value of $Q_A$ is of no importance
in the argument above.
The authors of \cite{Danielson}
take this to be
\begin{eqnarray}\label{DSW} 
Q_A \, = \, d^2 \, m_A, 
\end{eqnarray}
as might be envisaged by dimensional 
considerations.
In what follows, we will see
that generically $Q_A$ might be actually expected
to be quite larger than this,
and that the discrepancy might be of importance 
as a signature of the emission being nonclassical.

We proceed now with our analysis.
%
%%change0
%We assume
%that the energy-momentum densities $T^{ab}_{0(1, f)}$
%($a, b$ are spacetime indices)
%of each configuration 0, 1 and $f$ of Alice's system
%generate the field states $|\phi_0\rangle$, $|\phi_1\rangle$
%and $|\phi_f\rangle$,
%
We are working under the assumption that the gravitational field
requires a quantum description,
and the field
sourced by $A$ is entangled with the superposed positions. We assume then
that the energy--momentum densities $T^{ab}_0$,
$T^{ab}_1$, $T^{ab}_f$
($a, b$ are spacetime indices)
associated with
the configurations 0, 1, and $f$ of Alice's system
do quantum correlate with the gravitational field they generate,
described quantum-mechanically by the field states 
$|\phi_0\rangle$, $|\phi_1\rangle$,
and $|\phi_f\rangle$ respectively,
%%change1  
and we write the state of Alice's system ($A$ + fields)
as
\begin{eqnarray}\label{superposed} 
|\psi\rangle \, = \, \frac{1}{\sqrt{2}} \,
\big(|x_0\rangle |\phi_0\rangle  + |x_1\rangle |\phi_1\rangle \big),
\end{eqnarray}
at $t =0$,
and
\begin{eqnarray}\label{recombined}
|\psi_f\rangle \, = |x_f\rangle |\phi_f\rangle,
\end{eqnarray}
at $t = T_A$,
where $|x_0\rangle$, $|x_1\rangle$, 
and $|x_f\rangle$
are the states of $A$ corresponding
to configurations 0, 1, and $f$,
namely, with the center of mass of
$A$ at coordinates $x_0$, $x_1$, and $x_f$
where these are taken
in the center of mass frame of Alice's system
which we choose as our reference frame. 
As far as the two positions of $A$ are well
separated ($\approx$ no overlap between the states of $A$
describing each given position),
the matter states can be taken as orthogonal.
%
%%change0
%As for the field states,
%we have
%$\langle\phi_1|\phi_0\rangle = \langle 0|\phi_0 - \phi_1\rangle$,
%with
%$|\phi_0 - \phi_1\rangle$ the field state generated by
%$T^{ab}_0 - T^{ab}_1$.
%
As for the field states,
in the linearized quantum theory of gravity ($g_{ab} = \eta_{ab} + h_{ab}$
%%proof0
with $\eta_{ab}$ the Minkowski metric and perturbation $h_{ab}$ small)
%%proof1
which we find appropriate in our circumstances,
we assume we can proceed analogously to the electromagnetic case
(with energy--momentum tensors replacing currents, cf. \cite{Ford}).
In the latter case,
the overlap $\langle \varphi' | \varphi\rangle$
between two states $\varphi'$, $\varphi$ associated with
the currents $j'^a$, $j^a$ 
can be written as the overlap between the vacuum
and the state associated with the current $j^a - j'^a$
\cite{BreuerPetruccione, Mazzitellietal}. 
Here,
we have
$\langle\phi_1|\phi_0\rangle = \langle 0|\phi_0 - \phi_1\rangle$,
with
$|\phi_0 - \phi_1\rangle$ the field state generated by
$T^{ab}_0 - T^{ab}_1$.
%%change1
%
The differences $T^{ab}_0 - T^{ab}_f$, $T^{ab}_1 - T^{ab}_f$
are what is expected to produce radiation
(the entangling part coming from the difference
of the two, $T^{ab}_0 - T^{ab}_1$, cf. \cite{Belenchia1}).  
These differences might be generically expected to be 
quite significant in the region of superposition,
thus giving $|\phi_0\rangle$ and $|\phi_1\rangle$
nearly orthogonal. 
It is assumed, however, that the two components,
even if nearly orthogonal at $t=0$, can be
fully recombined in time $T_A$ \cite{Belenchia1}.
We will return to this later.

As for the emission 
during recombination,
we are free to probe the possibly emerging radiation where
we like. 
We use this freedom,
imagining to analyze it
at distances much larger than $d, {\bar x_A}$;
$d \ll D$ is also the condition used in 
\cite{Belenchia1, Belenchia2, Danielson}.
Assuming there is an analogy with classical emission,
this enables us to describe the radiation in terms of multipole expansion
in powers of $1/r$,
having taken $r$ the distance from the center of mass $C$
of Alice's system,
and in practice approximate it with the lowest-order terms.
What matters are the differences between the configurations
0 or 1 and $f$.
Assuming momentum (and angular momentum)
conservation during recombination,
the dipole term gives vanishing contribution,
and the emission is determined by 
(the third time derivatives of) quadrupole moments $Q$.
We have to consider the differences 
$Q(0) - Q(f)$, $Q(1) - Q(f)$, and $Q(0) - Q(1)$
(with obvious meaning of the notation),
with the latter being the part responsible for emission of entangling
radiation (in this case,
the field emitted is entangled with the source
and then brings decoherence and fading out of the final interference
pattern)~\cite{Belenchia1}.  
The latter quantity is also clearly responsible
for the difference of the two superposed fields
felt by a probe at $r$ (as, in particular, $B$).  
%%1

In view of our task of determining
the conditions for graviton emission, 
what we perform here is simply computing these differences 
of quadrupole moments,
having in mind a situation generic
with $A$ delocalized starting 
from a position not necessarily coinciding with $C$,
with coordinate $\bar x_A$ 
(which we take as non-negative) with respect to $C$ taken as origin.

%%0
We find
(the calculation is spelled out in the Appendix)
that, denoting $2 Q_A \equiv Q(0) - Q(1)$, 
$Q_A$ turns out indeed to be also
$Q_A = Q(0)- Q(f) = |Q(1) - Q(f)|$, 
at least when ${\bar x_A}$ is significantly larger than $d$,
and
\begin{eqnarray}\label{emit69.8}
Q_A = 2 m_A {\bar x_A} d = \frac{2 {\bar x_A}}{d} m_A d^2.
\end{eqnarray}
In particular,
a same $Q_A$ thus rules emission and entangling emission
(and which-path discrimination at distance $r$). 
%%1

As for the avoidance of the paradox
along the lines of \cite{Belenchia1, Belenchia2, Danielson},
this changes nothing
for it amounts merely to replace (\ref{DSW}) with (\ref{emit69.8})
in giving $Q_A$ (assuming $D \gg d,{\bar x_A}$).
It just confirms that we can use a same $Q_A$ in 
(\ref{emit55.5}) and (\ref{emit54.2}).

%%change0
%We see that (\ref{emit69.8}) gives 
%a same dependence on $m_A$ 
%and $d$ as (\ref{DSW}).}
%
We see that (\ref{emit69.8}) gives 
a dependence on the quantity $m_A d^2$, 
which is $Q_A$ in (\ref{DSW}),
with a factor $2 \bar x_A/d$ in front.
%%change1 
%
%%proof0
Both give, then, a threshold time for emission 
which goes like $\sqrt{m_A}$. 
%%proof1
%
The factor $2 {\bar x_A}/d$ in (\ref{emit69.8}) can give,
however, a value for $Q_A$ in principle 
much bigger than (\ref{DSW}), when ${\bar x_A}$ is significantly larger
than $d$, and this has consequences for the limit time.
%%0
From (\ref{emit54.2}) and (\ref{emit69.8}), we obtain
%%1
\begin{eqnarray}\label{emit69.9}
T_{\rm emit} 
\sim \sqrt{Q_A} 
= \sqrt{\frac{2 {\bar x_A}}{d}} \sqrt{m_A} \, d
= \sqrt{\frac{2 {\bar x_A}}{d}} \sqrt{\frac{m_A}{m_p}} \frac{d}{c},
\end{eqnarray}
where, in the last equality, we reinserted all constants with 
$m_p$ the Planck mass and $c$ the speed of light in vacuum.

%%0
In the first step here (which is Equation (\ref{emit54.2})),
we are making a classical analogy and taking $\frac{1}{T_A}$
as basic, characteristic scale of the angular frequency $\omega$
of the emerging radiation~\cite{Belenchia1}
(similar to if recombination were obtained harmonically
or similar to in the emission part of Thomson scattering).
The emitted power is $\sim$$({\dddot Q}_A)^2$
(dots are time derivatives) and 
the emitted energy $\cal E$ in time $T_A$ can be written
as
\begin{eqnarray}
{\cal E} 
\sim
\int_0^{T_A} {\dddot Q_A}^2 \, dt
= {\dddot Q_A}^2 \, T_A, 
\end{eqnarray}
with
${\dddot Q_A} \sim \frac{1}{T_A} \Big(\frac{1}{T_A} \, \frac{Q_A}{T_A}\Big)$.
If $\cal E$ is in gravitons of energy $\frac{1}{T_A}$,
the emission of at least one of them
is possible indeed only if $T_A < \sqrt{Q_A}$ \cite{Belenchia1}. 
Emission at lower frequencies can also be expected both in a classical
(Fourier transforming a generic pulse of duration $T_A$,
angular frequencies up to $\frac{1}{T_A}$ are significantly present)
and a quantum setting (population of low-energy states
in the radiation field),
even if depressed (radiated power is $\omega^6$
for a quadrupole source)
and of lower impact in reducing the coherence of $A$.  
This means that we have a tail of emission also when 
$T_A > \sqrt{Q_A}$,
but the gross picture is that there is an effective 
threshold recombination time
of a characteristic scale $\sim$$\sqrt{Q_A}$ to have emission;
this is what we refer to when considering the conditions
for the onset of emission.
%%1

%%0
This analysis is for when the gravitational field
does entangle with the superposed locations.
This corresponds
to having the field state at a point
as a superposition of (nearly orthogonal) states
$|\phi_0\rangle$  and $|\phi_1\rangle$, as said,
and gives $\approx$$\pm Q_A$
as variations of quadrupole moments
associated with each of the branches
of the superposition
and $\approx$$2 Q_A$ as difference 
of quadrupole moments between the branches.
In case the field does not entangle with the superposed locations 
and is sourced instead by the expectation value
of the energy--momentum tensor on the delocalized state,
we have
a single expression for the field $\phi$
given as the gravitational field sourced by 
the mass density distribution:
\begin{eqnarray}
\rho(x^i) 
&=&
\frac{1}{2} \,
\big[
m_A \delta(x^i - x^i_0) +
m_A \delta(x^i - x^i_1)\big]  
+ \rho_l
\end{eqnarray}
(describing $A$ in terms of Dirac's $\delta$)
where
$x^i_0 = (x_0, 0, 0)$
and $x^i_1 = (x_1, 0, 0)$
are the two superposed positions of $A$,
and
$\rho_l$ is the mass density distributions
of Alice's lab, $A$ excluded.
In this case,
we cannot have a difference connected with positions 0 and 1 
(energy--momentum densities of both positions contribute 
to the field, no which-path possible), 
and the variation $\widetilde Q_A$,
and also the emitted energy,
we obtain
in the recombination 
is much smaller than in the case
that the field is entangled.
Indeed, as can be easily verified (cf. Appendix),
$\widetilde Q_A = \frac{1}{2} d^2 m_A = 
\frac{1}{2} Q_A/(\frac{2 {\bar x_A}}{d})$ which is $\ll Q_A$
for $d \ll {\bar x_A}$.
We see that the factor $\frac{2 {\bar x_A}}{d}$ gives the order of magnitude
of $Q_A/\widetilde Q_A$, the gain we have 
in the amplitude if the field entangles,
and $(\frac{2 {\bar x_A}}{d})^2$ the gain in emitted energy.
%%1

The factor $d/c$ in (\ref{emit69.9}) clearly has the meaning
of absolute lower limit to the recombination time $T_A$
for two paths separated by a distance $d$: $T_A \ge d$ always
(cf. \cite{MariDePalmaGiovannetti}). 
%%0
In addition,
if the recombination takes places in a time $T_A$,
we have to consider that 
only a portion of size $c T_A$ (inserting, explicitly, the speed of light) 
of Alice's system can be involved in momentum transfer
(this being the part involved in overall momentum conservation 
during the recombination of $A$),
as components of Alice's system far from one another
and from $A$
by more than this distance
have no time to talk each other in reaction
to $A$ recombination.

This suggests that when following
the approach of \cite{Belenchia1, Belenchia2, Danielson},
one has to be careful with the definition of the system
under examination.
In particular, we have to consider the consequences of
that;
along with \cite{Belenchia1, Belenchia2, Danielson},
we {\it assume}
that $A$ can be coherently recombined in time $T_A$, 
namely, with the fields following the matter evolution
from the superposed state (\ref{superposed}),
with nearly orthogonal $|\phi_0\rangle$ and $|\phi_1\rangle$, 
to the recombined state (\ref{recombined}).
Indeed, the gravitational field at points farther than $c T_A$
from particle $A$ cannot be causally affected
by Alice system's evolution during recombination,
and then the assumption that the superposition
can be coherently recombined
requires, for consistency, that
only that same portion of size $\approx$$c T_A$ of the system  
is involved.
This would correspond to the view that the local physics
of $A$ and its neighborhood
ought to be ruled in terms of quantities
causally connected with it.  
Alice's system can thus be considered as 
effectively made by this portion with particle $A$ and the fields there,
and $C$ ought then to be taken as the center of mass of this reduced system.
This implies that we always have $T_A > 2 {\bar x_A}$ 
in Equation (\ref{emit69.9}).

The latter condition is the only remnant of this local physics request.
Looking at the Appendix, we see, indeed, that all the calculation
carries over for this restricted system provided the reduced 
system mass $M'$
is still $\gg m_A$, and gives the same results (\ref{emit64.1}) 
and (\ref{emit64.3}), with $Q_A$ given again by (\ref{emit69.8}). 

Concerning the gedanken experiment, 
for $T_A < D$ 
(which is the interval we are interested in), 
this, strictly speaking, leaves out $B$ and the region around it. 
This is not a real problem, however,
since, from the considerations just made,
the (differences of the) fields felt by $B$ from the restricted configuration
(fields suitably extended till reaching $B$)
are practically the same
as those actually sourced by the full mass distribution
of Alice's~system.

A tension between the request that Alice is able
to recombine $A$ in a finite time $T_A$ and the fact
that the fields do extend far beyond $c T_A$ 
and cannot keep up with the evolving source
was emphasized in \cite{Avilaetal}.
There, it was argued that this can be taken
as showing that, 
if $A$ can actually be coherently recombined in time $T_A$
and thus is able to provide interference patterns,
then the field states at $t=0$
ought not to be nearly orthogonal,
but instead have sizable overlap.
Building on this, Ref. \cite{Avilaetal} showed this would imply that the interference fringes,
if present in the absence of a test body
(this meaning the fields produce negligible decoherence),
would never disappear, whichever body might interact with the fields
(in particular, our test particle $B$).
This eliminates any possibility of superluminal communication
with $A$ and would solve, once and for all, 
any potentially paradoxical issue in the gedanken experiment.   
One might feel, however, a little uncomfortable with the fact that
in this approach,
the gravitational fields sourced by a delocalized (and then recombined) 
particle, 
such as, for example, those in the full-loop Stern--Gerlach interferometers
envisaged in the proposals \cite{Bose, Marletto},
apparently would be far from being orthogonal to each other
even if sourced by orthogonal states. 

Here, we take quite a complementary view.
On the basis of the fact that the physics of $A$
should also be describable in terms of only
local quantities,
namely, quantities able to affect $A$ in time $T_A$,
we maintain that if Alice's system (including its fields)
is effectively restricted to a causally connected neighborhood of $A$,
we can actually
have coherent recombination starting from orthogonal
$|\phi_0\rangle$, $|\phi_1\rangle$,
thus having orthogonal field states corresponding
to orthogonal source states. 
We stick, thus, to the protocol \cite{Belenchia1, Belenchia2, Danielson},
only noting that it requires a redefinition of the
system effectively taking part in the action.
This might not seem to solve all potential issues (cf. \cite{Avilaetal}) 
accompanying the gedanken experiment in the protocol~\cite{Belenchia1, Belenchia2, Danielson}, 
but shows at least the viability of coherent recombination
of orthogonal states in the mentioned restrictions.
This is all we need here after all,
our focus being not an examination 
of the gedanken experiment, 
but, instead, the possible graviton emission,
and the latter is dictated by the variations
$Q(0) - Q(f)$, $Q(1) - Q(f)$,
which do not depend, as discussed, on the approach.
%%1

This said, we search now for conditions which allow
for graviton emission. For this, we must have
\begin{eqnarray}\label{emit70.4}
d < T_A < \sqrt{Q_A},
\end{eqnarray}
which is
\begin{eqnarray}\label{emit70.5}
1 < \frac{T_A}{d} 
< \sqrt{\frac{m_A}{m_p}} 
\sqrt{ \frac{2 \bar x_A}{d} } ,
\end{eqnarray}
inserting explicitly the Planck mass.

Inequalities (\ref{emit70.5}) give
\begin{eqnarray}\label{emit70.6}
1 < \frac{T_A}{d} < \sqrt{\frac{m_A}{m_p}} \sqrt{ \frac{T_A}{d} },
\end{eqnarray}
%
%%proof0
which is clearly impossible to satisfy as long
as $m_A < m_p$, for $\frac{T_A}{d} > 1$ implies 
$\frac{T_A}{d} > \sqrt{\frac{T_A}{d}}$.
%%proof1
%
That is,
if $m_A < m_p$, we can never have $T_A < T_{\rm emit}$, 
i.e., graviton emission associated with recombination, 
and this is regardless of the choice of $\bar x_A$.  
The Planck mass acts as a lower-limit threshold mass $m_{\rm emit}$ 
for quadrupole emission, 
the latter being possible only if $m_A > m_{\rm emit} = m_p$. 

Present technology, and that foreseen in the near future,  
gives a delocalized $m_A \ll m_p$ by far.
Alice's (thought) experiment on $A$
(as well as the action on each
of the two delocalized
particles in actual experimental proposals \cite{Bose, Marletto}
checking for the nonclassical nature of gravity)   
is akin to completing
a Stern--Gerlach apparatus
with a recombination stage 
to obtain a proper Stern--Gerlach interferometer.
A first realization of such a device
was recently reported 
at single-atom level \cite{Folman},
with possible extensions of this same experimental procedure
to nanodiamonds ($10^6$ carbon atoms, $m_A \approx 10^{-20} \, {\rm kg}$) 
appearing within reach.
In addition,
for microdiamonds of $m_A \approx 10^{-14} \, {\rm kg}$
(radius $\approx$ 1 $\mu$m),
coherence times of $\gtrsim$1 s
might be conceivable under cooling \cite{Bar-Gill},
and
delocalizations of objects of this mass
with separation of order of their size 
might be within reach soon \cite{Pino}.
These figures are expected to be good enough
for proposals \cite{Bose, Marletto}
to start to be effective,
but, anyway, leave $m_A \ll m_p = 2.18 \cdot 10^{-8} \, {\rm kg}$.

Looking at present and near-future capabilities,
we thus cannot have graviton emission
associated with recombination even if $T_A$ is taken
as short as causally allowed.
However, from a theoretical point of view,
we can be allowed to imagine 
full-loop Stern--Gerlach interferometers
working with $m_A > m_p$.
In them,
$A$ can be recombined fast enough to allow
Alice's system to emit.
%
%%0
In (\ref{emit70.5}) (right inequality), we see that
the threshold time $T_{\rm emit}$ depends 
on $\bar x_A$.
The best option for allowing emission for a given $T_A$
is to have $\bar x_A$ as large as possible,
namely, such that $2 \, {\bar x_A} = T_A$.
We assume that this choice is not only possible if $T_A$ is just above $d$
but that it is generically realized with
Alice's system being macroscopic.
%%1
%
With it,
inequality (\ref{emit70.5}) coincides
with (\ref{emit70.6}) and allows that we have emission
when
\begin{eqnarray}\label{emit70.8}
T_A  < \frac{m_A}{m_p} \, d,
\end{eqnarray}
that is,
\begin{eqnarray}\label{emit_final}
T_{\rm emit} \sim \frac{m_A}{m_p} \, d,
\end{eqnarray}
which can also be derived from (\ref{emit69.9}),
taking $2 {\bar x_A} = T_{\rm emit}$.
We see that the condition of emission
depends this way generically on parameters concerning
particle $A$ alone ($m_A$, $d$)
as one might have hoped,
not on Alice's lab.
Notice that inequality (\ref{emit70.8}) gives
a threshold time which grows linearly with $m_A$,
not as $\sqrt{m_A}$, as might seem to be inferred
instead from~(\ref{emit69.9}).

%%proof0
This as far as the ability of Alice's system to emit is concerned. 
%%proof1
%
Regarding, instead, the paradox,
notice that 
its avoidance when $m_A < m_p$
is in that for these masses we cannot have (by far)
$\sqrt{Q_A} > D$, 
and thus Bob cannot perform which-path in $T_B < D$ in the first place.
Indeed, from~(\ref{emit69.8}) (with $m_p$ inserted),
we have
\begin{eqnarray}\label{emit79.5}
\sqrt{Q_A}/D = 
\sqrt{2 {\bar x_a}/D} \sqrt{m_A/m_p} \sqrt{d/D},
\end{eqnarray}
which clearly is $\ll 1$ for $m_A < m_p$
if $\bar x_A, d \ll D$.

Generic $m_A > m_p$ is still not enough 
for the potential onset of the paradox.
In view of~(\ref{emit79.5}),
we have to indeed require $m_A \gg m_p$ in order to have
$\sqrt{Q_A} > D$.
When $m_A$ is large enough to give this,
Alice's system necessarily emits~\cite{Belenchia1, Belenchia2, Danielson}, 
as described above,
and no paradox can in anyway arise.  

Inequality (\ref{emit70.8}) coincides
with the mentioned
minimum discrimination time reported in \cite{MariDePalmaGiovannetti}
(Equation (3) in \cite{MariDePalmaGiovannetti})
needed to avoid the paradox,
in spite of being 
(quite unconvincingly, cf. \cite{Belenchia1})  
derived there   
from consideration of dipole gravitational moments
(that is, neglecting 
the reaction of Alice's lab to 
the displacements of particle $A$,
a reaction which brings instead to momentum, 
and thus dipole moment, conservation);
notice, however, that 
according to our results,
the no-paradox argument used in \cite{MariDePalmaGiovannetti}
can be leveraged
only when $m_A \gg m_p$, as just mentioned.

Further,
if we imagine that
Alice checks the coherence of particle $A$
through an interference experiment
(as considered in \cite{BaymOzawa, Belenchia1, Rydving}),   
the minimum allowed time to have
the fringes ideally discernible 
(on account of the finite spatial resolution
limit $l_p$)
does coincide with the threshold time (\ref{emit_final}) 
for emission.
Indeed,
following \cite{Rydving},
if we call $\delta$ the fringe spacing,
we have (with all constants)
$
\delta \sim 
\lambda v T_A/d
\sim
l_p \frac{m_p}{m_A} \frac{c T_A}{d},
$
where $v$ is the velocity of $A$, 
$\lambda = h/(m_A v)$ its de Broglie wavelength, and
$h$ is (unreduced) Planck constant.
From this,
requiring that
$\delta > l_p$,
we obtain (\ref{emit70.8}).

This finding of
equivalence/coincidence between no-emission condition
and (ideal) detectability of fringe pattern in an interference
experiment
is at variance with \cite{Rydving},
where (using (\ref{emit54.2}) with $Q_A$ given by (\ref{DSW}))
the visibility of fringes 
is found to constrain more than no-emission
(when emission sets in, the fringes are undetectable already).
However, when $\bar x_A$ is not maximal 
(i.e., when $2 \, {\bar x_A} < T_A$, quite a nongeneric situation
as we mentioned)
we also find, as per \cite{Rydving}, that when emission sets in,
fringes' visibility is already lost.  
The general picture we obtain is that
emission has all that is needed to avoid the paradox,
but
fringes' discernibility taken alone (i.e., without considering emission)
is also fine for this;
moreover, in generic circumstances, the two requirements do coincide. 
They are then basically equivalent 
concerning the avoidance of the paradox
in an interferometric setup.

This confirms the stance \cite{Rydving} that, 
at least as far as checking of coherence 
of $A$ is carried out through interference,
the limit posed by existence of a limit length $l_p$ is enough
%%proof0
(without, strictly speaking, a need of bringing into play emission, 
%%proof1
but being, as we find here, equivalent
to the no-emission condition) 
to avoid any clash between complementarity and causality.
This is also what \cite{BaymOzawa} found
(though neglecting there, too,
the abovementioned reaction of Alice's lab to the displacements of $A$,
i.e., using dipole gravitational momenta).

In 
\cite{Danielson}, however, a different setting 
to probe the coherence of $A$ is considered,
not relying on the detection of an interference pattern.
Looking at this, it seems we have inevitably to require
graviton emission to avoid the paradox in case $m_A \gg m_p$
(clearly, provided gravity is supposed to be able to entangle;
if not, no paradox can arise).
%
%%proof0
This occurs when supposing that locality holds.
%%proof1
%
Assuming, instead, nonlocality 
of the gravitational communication channel
(as contemplated in \cite{FragkosKoppPikovski}),
it is not clear how to avoid the paradox (when $m_A \gg m_p$)
since
we have, of course, causality anyway
and no quantized mediators to react on $A$
which is causally disconnected from $B$;
but we will return to this in the final comments of the paper.
The consideration of the potential paradox 
might highlight
a possible weakness of (causal) nonlocality of the channel
as compared to locality.

%%%%%%%%%%%%%%%%%%%%%%%%%%%%%%%

\section{Contrasting with collapse models}

 In (\ref{emit69.9}) and (\ref{emit_final}),
the Planck mass $m_p$ plays a pivotal role
in that it sets the mass threshold for particle $A$ to emit. 
%
%%0
In particular, these expressions
say, as discussed, that if we have a delocalized particle
we cannot expect quadrupole emission on recombining it
if $m_A < m_p$. 
%%1

We would like to ask now how this compares
with Di{\' o}si's and Penrose's hypothesis~\cite{Diosi1, Diosi2, Penrose}
that any such superposition of a mass $m$ in two locations
is unstable when the mass is large,
and collapses or decays to one of the two locations 
with average lifetime $\tau = \hbar/E_{\Delta}$
(all constants in), where $E_{\Delta}$ is the
gravitational self-energy
of the difference of mass configurations 
%%proof0
in the two locations, up to a multiplicative 
%%proof1
constant~(\cite{HowlPenroseFuentes} for details, see also \cite{Diosi3}). 
%%0
%%proof0
As a matter of fact, this model seems ruled 
%%proof1
out in its basic formulation \cite{Donadietal},
but there is still a dependence on some parameters.
%%1
We ask
for which masses $m$ 
the decay time $\tau$
keeps being large enough to allow 
for quadrupole emission from recombination
if $T_A$ is taken sufficiently short.

%%0
For this, we take the expression 
$\tau = \frac{5}{6} \, {R}/{m^2}$
of \cite{HowlPenroseFuentes}
for a uniform massive delocalized sphere with radius $R$,
valid when the separation is $d \gg R$
and for a specific/reasonable choice of the multiplicative constant
(given by the parameter $\gamma$ in \cite{HowlPenroseFuentes}
set to $\frac{1}{8 \pi}$, which is $\sim$$50$ times smaller
than the value ruled out in \cite{Donadietal},
thus giving $\sim$$50$ times longer lifetimes).
%%1
%
The exact expression of $E_{\Delta}$ grows rapidly at increasing $d$
from 0
at $d = 0$ to being already roughly 2/3 of the value quoted above
at $d = 2 \, R$ \cite{HowlPenroseFuentes}.

This clearly gives an upper limit $\tilde m$ to mass
to leave $\tau$ large enough for the above.
This can easily be estimated as follows.
If we take the separation as short as $d = 2 \, R$,
corresponding to have the two superposed mass distributions
on the verge of overlapping,
we must have
\begin{eqnarray}
\nonumber
2 \, R = d < \frac{3}{2} \cdot \frac{5}{6} \, \frac{R}{m^2},
\end{eqnarray}
which gives
$m < {\tilde m} = \sqrt{5/8} = 0.79 \, m_p$,
inserting, explicitly, the Planck mass.
Any larger $d$ at mass fixed means a larger recombination time
and, in addition, a smaller $\tau$;
for any given mass, the best option to obtain  
recombination time $< \tau$ is then to choose $d = 2R$,
and the just-given $\tilde m$ is the largest allowed mass
to have this inequality satisfied.  

There is clearly a tension between the collapse
models on one side and the possibility to obtain quadrupole emission
from recombination on the other.
When $m_A$ is, indeed, large enough to allow, in principle,
for emission ($m_A > m_{\rm emit} = m_p$), the collapse models foresee
it to decay before it can recombine
(and, if we read this the other way around,
the delocalization itself of such an $m_A$ is problematic
in the first place, with $m_A \approx m_p$ playing, then, the role
of an upper-limit
mass scale for delocalization to possibly happen in collapse models,
cf. \cite{ChristRovelli}).
If the proposal of Di{\' o}si and Penrose 
(in its basic form) 
is correct,
there is no possibility to obtain (quadrupole) emission
%%proof0
while recombining $A$; this, whichever $m_A$ is and
%%proof1
however small we 
(consistently) take 
the recombination time $T_A$. 

Something that is a little bit striking
is the coincidence $m_{\rm emit} \approx~{\tilde m}$
between the (lower-limit) threshold mass $m_{\rm emit}$ for quadrupole emission
from recombination and the (upper-limit) threshold mass $\tilde m$ 
to have the collapse proposal allowing for the delocalized particle
to have enough time to recombine (and have it delocalized
in the first place).
%
%%proof0
Things happen as if when circumstances 
would finally allow for emission
(delocalized masses large enough), 
right then the latter is inhibited by the collapse. 
%%proof1

As for the paradox,
one thing that the consideration of collapse models adds
is that if Di{\' o}si and Penrose are right,
the crucial case $m_A \gg m_p$ can never happen.
This immediately means that no paradox can arise
(in particular, no need to invoke graviton emission),
and, in addition, looking at (\ref{emit79.5}) and (\ref{emit55.5}),
that long integration times are needed for $B$ to possibly obtain which-path
(this being hardly compatible with a noncollapsing $A$ \cite{Aspelmeyer}).

%%%%%%%%%%%%%%%%%%%%%%%%%%%%%%%

\section{Summary and conclusions}

We tried to determine the conditions
for graviton emission from recombination
of a delocalized particle.
This was carried out having as background
the gedanken experiment~\cite{MariDePalmaGiovannetti, Belenchia1}
(in which Alice recombines a delocalized particle ($A$)
while 
Bob tries to perform which-path 
a distance $D$ apart
with a test particle ($B$);
in this, a tension between causality and complementarity
might potentially arise when Alice and Bob act in times $T_A, T_B < D$
if we assume that the gravitational field sourced by $A$
entangles with the
superposed~locations).  

To this aim,
we simply explicitly computed, for generic geometric conditions, 
the variation of quadrupole moments 
(of the delocalized particle and its lab, which we called Alice's system) 
from before to after $A$' recombination,
both in case the field is entangled with the positions
and in case it is not and is instead sourced by the expectation
value of energy--momentum on the delocalized state.
In view of the gedanken experiment, we also computed
the difference between the quadrupole moments
of the superposed~configurations.

We found that the variation of the moments in the recombination
is greatly enhanced in case the field is entangled compared to if it is
instead sourced by the energy--momentum on the delocalized state
(in which case the variation is simply $\sim$$m_A d^2$,
i.e., what is naively expected on dimensional grounds)
and provided the gain.  
We provided a formula for how quickly 
recombination must occur
for graviton emission to set in.
In it, the threshold time for graviton emission
grows as $m_A$ in place of $\sqrt{m_A}$ (which is 
what is obtained instead if the variation of quadrupole moment 
is $\sim$$m_A d^2$).    
In all this,
graviton emission is found to be possible
only when $m_A > m_p$ for recombination times short enough,
meaning that for masses smaller than the Planck mass,
no graviton emission is possible, however small we (consistently)
take the recombination time.

Concerning the gedanken experiment,
from the computed difference of the moments 
in the superposed configurations,
we find that a potential clash between causality
and complementarity is, in principle, conceivable
only when $m_A \gg m_p$
(which comes from requiring Bob to be able to perform which-path in $T_B < D$). 
Clearly, no clash can arise, however, since for these masses,
if $T_A < D$, Alice's system necessarily emits,
%%proof0
and the coherence of $A$ is affected without need of
%%proof1
causal relationship with $B$,
along the lines of \cite{Belenchia1, Belenchia2, Danielson}.
If the coherence of $A$ is probed, in particular,
through inspection of
interference fringes when $A$ is recombined,
the condition for the onset of emission 
turns out to coincide with the condition
of the separation $\delta$ of the fringes
to become $\delta < l_p$,
so that the two conditions
of the onset of emission on one side 
and the disappearing of the interference pattern (at ideal conditions)
on the other,   
do result as equivalent in this setting.
If, instead,
the probing of the coherence
%%proof0
of $A$ is performed in another manner (as proposed in \cite{Danielson})
not relying on the detection of the interference pattern,
%%proof1
it seems crucial that graviton emission sets in
to avoid any clash between causality and complementarity. 

This brings with it that
if the communication channel is assumed to be nonlocal---instead of local, as implicit in discussion above concerning emission---yet causal,
as contemplated (together with the local channel) 
in \cite{FragkosKoppPikovski},
it is not so clear how to avoid the paradox when $m_A \gg m_p$
in the noninterferometric setting of \cite{Danielson},
since we do not have interferometric fringes
to wash off (with finite limit $l_p$), 
nor we can rely on emission  
for having $A$ to decohere while recombining it in $T_A < D$;
yet performing which-path of $B$ is ideally possible within $T_B < D$,
this potentially clashing with complementarity.

When all this is considered within the collapse models 
of Di{\' o}si and Penrose \cite{Diosi1, Diosi2, Penrose}
(in their basic formulation),
we saw
that the case $m_A \gg m_p$ can never happen
(since the delocalized state decays before it recombines,
or before it can be formed in the first place),
and then no paradox can arise since Bob will never be able
to perform which-path in $T_B < D$.

Indeed,
in these models
it is not possible 
to have $A$ delocalized, 
even when, simply, $m_A > m_p$.  
Connected to the above, this means that
(quadrupole) emission from recombination 
would be never possible in them.
More precisely, we have the curious coincidence
$m_{\rm emit} \approx \tilde m \, \, (\approx$$m_p)$
between the threshold mass $m_{\rm emit}$ for emission 
and the threshold mass $\tilde m$ to have separation
that withstands decay,
meaning that
right when $m_A$ would be large enough  
to obtain emission (with $A$ recombining in a time
as short as possible),
it would then become also too large
to have $A$ not collapsed yet in one of the two locations.

In closing,
we would like to make a comment on the role
of Planck length $l_p$ in the above.
We saw that
the onset of graviton emission in the recombination 
of a delocalized particle
and the washing out of the pattern
in an interferometric setting due to the limit $l_p$
are two sides of the same coin.
This may lead one to suspect that 
the existence of a limit length alone,
when suitably introduced in the formalism,
might account for a great deal of results
concerning quantum features of curvature (cf. \cite{Nicolini}),
this clearly
irrespective of the actual underlying quantum theory of gravity.

The systematic investigation of all the consequences
of a limit length is the goal of
the framework \cite{KotE, Pad01}
(called minimum-length or zero-point-length metric or qmetric), 
which computes the distance
between two points, $p$ and $P$, with 
a lower-limit-length built in,
thus with smallest-scale nonlocality 
embodied in the biscalar, which provides distances.
In this, tensors are replaced by bitensors
as fundamental objects in the description,
with some selected ones playing a major role.
In particular, the metric tensor is replaced by
a (qmetric) bitensor which, consistent with the need to provide
a finite limit length, diverges in the coincidence limit. 
We might speculate that the loss of coherence of particle $A$,
as described here
at weak-gravity conditions (we use Newtonian gravity),
might be reobtained as an effect of
the qmetric associated with Minkowski 
(that is, replacing Minkowski  
with qmetric Minkowski),
regardless of graviton emission;  
a hint, in this sense, might be 
that, assuming gravity has mediators, 
the threshold mass for graviton emission 
we obtained turns out to be the Planck mass
(this would cure, by the way, the problem mentioned above
of how to avoid a clash between causality and complementary
in case we lack interferometric fringes to wash off
or gravitons to emit).  

On a parallel side,
some intriguing curvature-related quantum effects
investigated through the use of key bitensors 
are discussed in \cite{HariKKothawala, SinghKothawala}.
%
%%proof0
In the qmetric,
a number of results have been obtained relating curvature, 
and the dynamics (field equations),
to an underlying quantum structure of spacetime
(see \cite{Pad20, PaddySumanta}); attempts to investigate the latter are 
detailed
in \cite{PesQ, PesR, PesT}.

{\it Acknowledgments.} I'm grateful to Alessio Belenchia 
for useful comments on an earlier version of the paper. 
This work was supported in part by INFN grant FLAG.

\vspace{1 cm}

%%%%%%%%%%%%%%%%%%%%%%%%%%%%%%%%%%%%%%%%%%%%%%%%%%
\section*{\large Appendix: Evaluation of gravitational gradients
and their variations}

We address the problem of determining the difference
of the gravitational gradients felt at a distance
in the two configurations 
corresponding 
to the two superposed positions 0 and 1 of $A$ (Figure \ref{fig}).
No dipole term can contribute to this difference \cite{Belenchia1};
the dipole term taken with respect to the center of mass of Alice's system
($A$ + lab of Alice)
is actually vanishing in any configuration.
In the circumstances assumed in \cite{Rydving}
(centers of mass of $A$ and of the lab of Alice
coinciding for undelocalized $A$), the quadrupole moments
in the two configurations are equal
and cannot affect the difference.
We claim here that if we consider the slight generalization 
of not-coinciding centers of mass,
the quadrupole moments with respect to the center
of mass of Alice's system are different in the two cases,
and they become the dominant contribution,
as in \cite{Belenchia1}.
Moreover, their variations result as much bigger
in case the field is entangled with the superposed locations
than if it is instead sourced by the expectation value
of the energy--momentum tensor on the delocalized state. 

To see how this comes about,
let us write the gravitational potential $\phi$
at a point of spatial coordinates $x^i$, $i = 1, 2, 3$
with respect to some origin,
%%proof0 
as (cf. e.g. \cite{MTW})
%%proof1
%
\begin{eqnarray}\label{35.1}
\phi = 
- \,
\Big(\frac{M}{r} + \frac{d_j n^j}{r^2} +
\frac{Q_{ij} n^i n^j}{2 r^3} + \cdots \Big),
\end{eqnarray}
where
$M$ is the mass of the body which is the source of the potential
(in our case, Alice's system: $A$ + lab of Alice),
$n^i = x^i/r$ with $r$ the distance to the origin,
$d^i$ is dipole moment,
and the quadrupole is
\begin{eqnarray}
Q_{ij} =
\int (3 x'_i x'_j - r'^2 \delta_{ij}) \, \rho \, dV,
\end{eqnarray}
where the integral runs over the body,
with $r'$ the distance to the point with the attached 
volume element $dV$ at coordinates $x'^i$, %Please ensure intended meaning is retained
and $\rho$ the density there.

We decide to compute $\phi$, taking as origin the center of mass $C$
of Alice's system. Clearly, this implies that $d^i = 0$, $i=1,2,3$. 
Now, at points along the $x$ axis, taken as the direction
connecting the superposed positions, 
we have
\begin{eqnarray}\label{35.3}
Q_{ij} n^i n^j =
\int (3 {x'}^2 - {r'}^2) \, \rho \, dV
\equiv
Q_{xx}.
\end{eqnarray} 
%
%%0
If we consider the approximation of
a mass distribution $\rho_A$ of the $A$ particle
given by a Dirac's $\delta$
(for the sake of simplicity, but this can be relaxed),
and 
assume that the particle has coordinate
$x_A$ with respect to $C$,
we obtain
\begin{eqnarray}\label{35.4}
\int (3 {x'}^2 - {r'}^2) \, \rho_A \, dV
&=&
\int (3 {x'}^2 - {x'}^2) \, m_A \, \delta(x' - x_A) \, dx'
\nonumber \\
&=&
2 \, x_A^2 \, m_A.
\end{eqnarray}
%
%%1

Considering the mass distribution of
the lab of Alice (meant specifically as the system
of Alice with $A$ removed),
and calling $x_{labA}$ the $x$-coordinate of its center of mass
with respect to $C$, 
we have
\begin{widetext}
\begin{eqnarray}\label{35.5}
\int (3 {x'}^2 - {r'}^2) \, \rho_{labA} \, dV
&=&
\int (3 {x'}^2 - {x'}^2) \, dM_A
\nonumber \\
&=&
\int 2 \, (x' - x_{labA} + x_{labA})^2 \, dM_A
\nonumber \\
&=&
2 \,
\Big[\int (x' - x_{labA})^2 \, dM_A + \eta \, x_A^2 \, m_A\Big], 
\end{eqnarray}
\end{widetext}
with
$M_A$ the mass of the lab, 
$\eta \equiv m_A/M_A$,
and, of course, $x_{labA} = - \eta \, x_A$.

Let us consider general circumstances 
in which the position of $A$ when not delocalized
is not coinciding with $C$
but has, instead, a slight offset $\bar x_A \ll r$
along
the $x$-axis   
($\bar x_A = 0$ in the circumstances of \cite{Rydving}).
In the configuration of Alice's system corresponding
to the particle $A$ in path 0 (see Figure \ref{fig}),
we have 
$
x_A^{(0)} = \bar x_A + d/2,
$
where the index $^{(0)}$ tags the configuration.
Analogously,
$
x_A^{(1)} = \bar x_A - d/2.
$

Calling $Q_{xx}(0)$ and $Q_{xx}(1)$ the corresponding quadrupoles, 
from (\ref{35.3}) we obtain
\begin{widetext}
\begin{eqnarray}
Q_{xx}(0) =
2 \, \Big({\bar x_A} + \frac{d}{2}\Big)^2 \, m_A +
2 \, \Big[\int (x' - x_{labA}^{(0)})^2 \, dM_A + 
\eta \, \Big({\bar x_A} + \frac{d}{2}\Big)^2 \, m_A\Big]
\end{eqnarray}
and
\begin{eqnarray}
Q_{xx}(1) =
2 \, \Big({\bar x_A} - \frac{d}{2}\Big)^2 \, m_A +
2 \, \Big[\int (x' - x_{labA}^{(1)})^2 \, dM_A + 
\eta \, \Big({\bar x_A} - \frac{d}{2}\Big)^2 \, m_A\Big]
\end{eqnarray}
\end{widetext}
with $x_{labA}^{(0)} = -\eta \, x_A^{(0)}$
and $x_{labA}^{(1)} = -\eta \, x_A^{(1)}$.
The two integrals here depend only on the {\it form} of mass
distribution of the lab around its actual center of mass in the two
configurations;
their difference,
as well as their difference with respect to the value $Q_{xx}(f)$
for the final configuration with particle $A$ recombined 
(i.e., $x_A = {\bar x_A}$), 
can be estimated 
to be ${\cal O}((\eta d)^2 M_A) = {\cal O}(\eta \, d^2 \, m_A)$
(and is identically vanishing in the approximation
of rigid displacement).

Considering the case of field entangled with the superposed positions,
we have that the variations
of the quadrupole moments in recombination
are $|Q_{xx}(0) - Q_{xx}(f)|$ and $|Q_{xx}(1) - Q_{xx}(f)|$
for positions 0 and 1, respectively,
while their difference in the two positions is
$Q_{xx}(0) - Q_{xx}(1)$. 
Neglecting terms containing $\eta$ as a factor, namely,
of order ${\cal O}(\eta {\bar x_A^2}, \eta {\bar x_A} d, \eta d^2)$,  
we obtain
\begin{eqnarray}\label{emit64.1}
Q_{xx}(0) - Q_{xx}(f) &=& 
2 \, \Big({\bar x_A} + \frac{d}{2}\Big)^2 m_A - 2 \, {\bar x_A}^2 m_A 
\nonumber \\
&=&
2 \, \Big({\bar x_A} d + \frac{d^2}{4}\Big) \, m_A
\nonumber \\
&=&
Q_A + \frac{d^2}{2} \, m_A,
\end{eqnarray}
with
$Q_A = 2 \, {\bar x_A} d \, \, m_A$,
and analogously,
\begin{eqnarray}\label{emit64.3}
Q_{xx}(1) - Q_{xx}(f) =
- Q_A + \frac{d^2}{2} \, m_A.
\end{eqnarray}

We have, then,
$Q_{xx}(0) - Q_{xx}(1) = 2 \, Q_A$
and, 
when ${\bar x_A}$ is significantly larger
than $d$ (though still with ${\bar x_A} \ll r$),
$|Q_{xx}(0) - Q_{xx}(f)| \simeq 
|Q_{xx}(1) - Q_{xx}(f)| \simeq Q_A$.
This proves Equation (\ref{emit69.8})
and what we said about its meaning in the main text.

We see that,
contrary to the case considered in \cite{Rydving},  
in general, the quadrupoles 
corresponding to the two configurations
are not equal, with a difference 
${\cal O}\big({\bar x_A} d \, \, m_A\big)$,
which provides
the dominant contribution to the difference $\Delta \phi$
in the gravitational field felt by $B$
(of course, in case of ability of the gravitational field to entangle
$A$ with $B$).

With this,
we can proceed to compute the difference $\Delta x$
in the position of particle $B$ associated with this $\Delta \phi$,
assuming $d, {\bar x_A} \ll D$.
We have
$
\Delta \phi = Q_A/r^3 \approx Q_A/D^3, 
$
and then
$
\Delta g = 3 \, Q_A/r^4 \approx 3 \, Q_A/D^4,
$
with $g$ denoting the acceleration of $B$.
We then have
$
\Delta x = 1/2 \,\, \Delta g \, T_B^2 = 3/2 \, Q_A/D^4 \, \, T_B^2
\sim Q_A/D^4 \, \, T_B^2,   
$
which, on imposing $\Delta x > 1$,
gives Equation (\ref{emit55.5}) 
in agreement with \cite{Belenchia1},
but with $Q_A$, as in Equation (\ref{emit69.8}).

If the field is not entangled with the positions
but is sourced instead by the expectation value of the 
energy--momentum tensor
on the delocalized state,
the quadrupole term is not a superposition of two terms
but has a well-defined value,
and no discrimination is possible between the paths.
%
%%0
There is, anyway, a variation $\widetilde Q_A$ of the quadrupole moment
in the recombination which can be calculated as
$\widetilde Q_A = \frac{1}{2} [Q_{xx}(0) + Q_{xx}(1)] - Q_{xx}(f) = 
\frac{1}{2} d^2 m_A$,
where $Q_{xx}(0)$ and $Q_{xx}(1)$ 
are as given in (\ref{emit64.1}) and (\ref{emit64.3}).
%%1
%
We see this is the same order of magnitude
of the naive guess (\ref{DSW})
and is, in general, much smaller than $Q_A = 2 \, {\bar x_A} d \, \, m_A$,
which is what we obtain instead if the field does entangle
with the positions.

%%%%%%%%%%%%%%%%%%%%%%%%%%%%%%%%%%%%%%%%%%%%%%%%%%

\end{document}